\documentclass[epj,final]{svjour}

\usepackage{latexsym}
\usepackage{url}
\usepackage{amsfonts}
\usepackage{amsmath, amssymb}
\RequirePackage{graphicx}

\begin{document}
\title{Lense-Thirring precession and gravito-gyromagnetic ratio}
\author{A.~Stepanian\inst{1}, Sh.~Khlghatyan\inst{1}, V.G. Gurzadyan\inst{1,2}
}                     
%
%
\institute{Center for Cosmology and Astrophysics, Alikhanian National Laboratory and Yerevan State University, Yerevan, Armenia \and
SIA, Sapienza Universita di Roma, Rome,  Italy}
\date{Received: date / Revised version: date}
%

\abstract{The geodesics of bound spherical orbits i.e. of orbits performing Lense-Thirring precession, are obtained in the case of the $\Lambda$-term within gravito-electromagnetic formalism. It is shown that the presence of the $\Lambda$-term in the equations of gravity leads to both relativistic and non-relativistic corrections in the equations of motion. The contribution of the $\Lambda$-term in the Lense-Thirring precession is interpreted as an additional relativistic correction and the gravito-gyromagnetic ratio is defined.} 
%
\maketitle

\section{Introduction}

The Lense-Thirring (LT) effect is one of essential predictions of General Relativity (GR) \cite{C1,C2} which enabled accurate measurements in conditions of the Earth gravity by means of laser ranging satellites \cite{C3,C4,C5}. The potential observability of L-T 
precession is among the features attributed to the accretion disks of galactic nuclei and binary stars, e.g. \cite{Dyda}.

LT precession has been efficiently treated within the gravito-electromagnetism (GEM) (or gravitomagnetism) formalism, see \cite{GEM,GEM1,GEM2,M1,M2} and refs therein. This approach will be used below, as enabling to reveal explicitly the contributions of the relevant terms in the equations of motion.  

We will be interested in the role of the $\Lambda$-term in the LT effect. This issue has to be considered within the importance of consideration of modified gravity models to describe the observational data on the dark energy and dark matter. Namely, we will use the following metric for the spherically symmetric solution of field equations 
\begin {equation} \label {mod}
g_{00} = 1 - \frac{2 G m}{c^2r} - \frac{\Lambda r^2}{3}\,; 
\qquad g_{rr} = \left(1 - \frac{2 G m}{c^2r} - \frac {\Lambda r^2}{3}\right)^{-1}\,,
\end {equation} 
known as Schwarzschild-de Sitter metric \cite{Rind}. This metric is arising also as weak-field limit of GR in view of the Newton's theorem on the “sphere-point” equivalency \cite{G,GS1,GS2,GKS}. Then, the cosmological constant, $\Lambda$ can be attributed as a fundamental constant \cite{GS3}, which is present in both relativistic and non-relativistic equations of gravity. Within McCrea-Milne cosmology it enables one to consider e.g. the observed galactic flow in the vicinity of the Local Group  \cite{RG,GS4}.

The appearance of $\Lambda$ in the above relations can be given a clear group-theoretical background. Namely, for three different vacuum solutions for GR equations the isometry groups are defined depending on the sign of $\Lambda$,
the stabilizer group of the maximally symmetric Lorentzian 4D-geometries is the Lorentz group O(1,3). Then, for all of these Lorentzian geometries the group O(1,3) of orthogonal transformations defines a spherical symmetry in Lorentzian sense at each point; for details see \cite{GS1}.

The numerical value of $\Lambda$ is too small i.e. $\Lambda = 1.11 \times 10^{-52}$ $m^{-2}$\cite{Pl} and certainly the possibility to observe/detect the role of the $\Lambda$-term in Eq.(1) looks non-trivial for now, as previously had appeared so, but now directly observed the black hole's shadow in the center of M87 galaxy and of the gravitational waves. Particularly, the  gravitational lensing can provide one of the possibilities to detect a discrepancy between GR and $\Lambda$-gravity \cite{GSlens}. Up to now, different constraints for $\Lambda$ are obtained \cite{Con1,Con2,Con3,L,SK}, with no inconsistency with the above mentioned numerical value for the cosmological constant.

In this paper, for the first time we merge the LT effect, GEM and the $\Lambda$-gravity.  We start with the geodesic equations in a time-varying LT system based on \cite{M1,M2}, and then move to the interpretation of the nature of additional $\Lambda$-term appeared in the LT precession in the context of GEM.

\section{Spherical orbits}

In the context of $\Lambda$-gravity, the metric of time-depending angular momentum of a central mass is given by following relation
\begin{equation}\label{GEMmetric}
ds^2=-c^2 \left( 1-2\frac{\Phi}{c^2} \right) dt^2-\frac{4}{c}(\mathbf{A}\cdot d\mathbf{x}) dt+\left( 1+2\frac{\Phi}{c^2} \right) \delta_{ij}dx^idx^j,
\end{equation}
where in the same analogy with standard GEM 
\begin{equation}
\Phi = \frac{GM}{r} + \frac{\Lambda c^2r^2}{6},\quad \mathbf{A}=\frac{G}{c} \frac{\mathbf{J}(t)\times \mathbf{x}}{r^3}, \quad 
\end{equation}
are the gravitoelectric and gravitomagnetic potentials, respectively. Meantime, $r = |\mathbf x|$ and $\mathbf{J}(t)$ linear time-dependent angular momentum of central mass
\begin{equation*}
\mathbf{J}(t)=(J_0+J_1t)\hat{\mathbf{z}}.
\end{equation*}   
Considering the geodesics $\frac{d u^\mu}{d \tau} +\Gamma^\mu{}_{\rho\sigma}u^\rho u^\sigma=0$, we will have
\begin{equation*}
\begin{aligned}
c^2\Gamma^0{}_{0\mu}&=-\Phi_{,\mu},\\ c^2\Gamma^0{}_{ij}&=2A_{(i,j)}+\delta_{ij}\Phi_{,0},\\
c^2\Gamma^i{}_{00}&=-\Phi_{,i}-2A_{i,0},\\ c^2\Gamma^i{}_{0j}&=\delta_{ij}\Phi_{,0}+2 A_{[i,j]},\\
c^2\Gamma^i{}_{jk}&= \delta_{ij}\Phi_{,k}+\delta_{ik}\Phi_{,j} -\delta_{jk}\Phi_{,i}.
\end{aligned}
\end{equation*}
From GEM point of view, the equation of motion can be regarded as the equation of motion for a charge due to the ``Lorentz force". It will take the following form
\begin{equation} \label{mot}
\frac{d {\mathbf v}}{d t}+\frac{GM{\mathbf x}}{r^3} - \frac{\Lambda c^2 {\mathbf x}}{3}= \left(\frac{GM}{c^2r^3} - \frac{\Lambda}{3}\right)[4({\mathbf x}\cdot {\mathbf v}){\mathbf v}-v^2{\mathbf x}]+\frac{2G}{c^2}\frac{\dot {\mathbf J}\times {\mathbf x}}{r^3}-\frac{2}{c}{\mathbf v}\times {\mathbf B} -\frac{6GJ(t)}{c^4r^5}[\hat {\mathbf J}\cdot ({\mathbf x}\times {\mathbf v})]({\mathbf x}\cdot {\mathbf v}){\mathbf v},
\end{equation}
where
\begin{equation*}
\mathbf B = \mathbf \nabla \times \mathbf A = \frac{G(J_0+J_1t)}{cr^5} (3z\mathbf{x} -r^2\hat{\mathbf{z}}).
\end{equation*}

The equation of motion includes the linear post-Newtonian contributions. However, in order to fully understand the motion of the test particle we have to include the following non-linear gravitoelectric term $\frac{4G^2M^2}{c^2r^4} \mathbf{x}$ too.

In spherical coordinates $(r,\theta, \phi)$ the Eq.(\ref{mot}) will be written as
\begin{align}
&\ddot{r}-r\dot{\theta}^2 -r\dot{\phi}^2\sin ^2 \theta +\frac{GM}{r^2} - \frac{\Lambda c^2 r}{3} =\frac{GM}{c^2r^2} \left( 4\dot{r}^2-v^2+\frac{4GM}{r}\right) - \frac{\Lambda r}{3} (4\dot{r}^2-v^2)
+\frac{2GJ(t)}{c^2r^2} \dot{\phi} \sin ^2\theta , \label{sph1}\\ 
&r\ddot{\theta} +2\dot{r}\dot{\theta} -r\dot{\phi}^2\sin \theta \cos \theta  = \left(\frac{4GM}{c^2r} - \frac{4 \Lambda r^2}{3}\right)   \dot{r}\dot{\theta} -\frac{4GJ(t)}{c^2r^2} \dot{\phi} \sin \theta \cos \theta , \label{sph2}\\ 
&\nonumber r\ddot{\phi}\sin \theta +2\dot{r}\dot{\phi} \sin \theta  +2r \dot{\theta} \dot{\phi} \cos \theta =\left(\frac{4GM}{c^2r} - \frac{4 \Lambda r^2}{3}\right) \dot{r}\dot{\phi}\sin \theta   
+\frac{2G\dot{J}}{c^2r^2}\sin \theta -\frac{2GJ(t)}{c^2r^3} (\dot{r}\sin \theta -2r\dot{\theta}\cos \theta ) \label{sph}.\\ 
\end{align}
The Eq.(\ref{sph}) can be rewritten as
\begin{equation}\label{int}
\frac{d}{dt} \left[ r^2\dot{\phi} \sin ^2\theta  -\frac{2GJ(t)}{c^2r} \sin ^2\theta \right] = \left(\frac{4GM r}{c^2} - \frac{4\Lambda r^4}{3}\right) \dot{r}\dot{\phi}\sin ^2\theta.
\end{equation}
Consequently, we get the integral of the motion if the right-hand side of Eq.(\ref{int}) vanishes. Considering spherical orbits $(\varphi, \vartheta, \rho)$, i.e. the circular orbits with frame dragging, we obtain
\begin{equation}\label{phi}
\dot{\varphi} =\frac{C}{\sin ^2\vartheta} +\frac{2GJ(t)}{c^2\rho^3}.
\end{equation}
Therefore, the Eqs.(\ref{sph1}, \ref{sph2}) will be written as
\begin{align}
\dot{\vartheta} ^2+\frac{C^2}{\sin^2\vartheta}&=\frac{GM}{\rho^3} \left( 1-\frac{3GM}{c^2\rho} - \frac{\Lambda \rho^2}{3}\right) -\frac{6GJ(t)}{c^2\rho^3} C - \frac{\Lambda c^2}{3} \left(1 + \frac{GM}{c^2 \rho} - \frac{\Lambda \rho^2}{3} \right)\\
\ddot{\vartheta} -\frac{C^2\cos \vartheta}{\sin ^3\vartheta} &=0.
\end{align}
These equations will be compatible only if $\dot{J}=0$. Thus by setting $J=J_0$, we define the positive constant $\Omega$ such that 
\begin{equation}\label{Om2}
\Omega^2 =\frac{GM}{\rho^3} \left( 1-\frac{3GM}{c^2\rho} - \frac{\Lambda \rho^2}{3}\right) -\frac{6GJ_0}{c^2\rho^3} C - \frac{\Lambda c^2}{3} \left(1 + \frac{GM}{c^2 \rho} - \frac{\Lambda \rho^2}{3} \right).
\end{equation}
It should be noticed that due to the incorporation of $\Lambda$ in the equations of gravity, according to Eq.(\ref{mod}), changes/modifies the notion of $\Omega$ which was introduced in \cite{M1}. Namely, as stated above we have both relativistic as well as non-relativistic corrections which are appeared due to the presence of $\Lambda$ term. First, according to Eq.(\ref{mod}), the standard Newtonian dynamics is written as
\begin{equation}\label{Kepler}
v^2 = \frac{GM}{r} - \frac{\Lambda c^2 r^2}{3}.
\end{equation}
The second corrections is pure relativistic which can be observed in Eq.(\ref{mot}). Indeed, in contrast to the left-hand side of Eq.(\ref{mot}) the $\frac{\Lambda}{3}$ term on the right-hand side is a pure relativistic effect and has no classical analogue.  

In this sense, although we have newly modified $\Omega$ term, comparing to the results of \cite{M1} the nature of the solutions do not change. Thus, for the motion in $\vartheta$ we get
\begin{equation}\label{solt}
\left( \frac{d\cos \vartheta}{dt}\right)^2 =(\Omega^2-C^2)-\Omega^2\cos ^2\vartheta.
\end{equation}
The above equation has solution once $\Omega^2\geq C^2$. Thus, we have
\begin{equation}\label{solcond}
\cos \vartheta =\alpha \sin (\Omega t+\beta), \quad C^2=\Omega^2 (1-\alpha^2),
\end{equation}
where $\beta$ is a constant. Accordingly, in the same analogy with \cite{M1}, if we take $C=\Omega \cos i$, the solution of the Eq.(\ref{phi}) may be written as
\begin{equation}\label{sol}
\varphi (t)=\frac{2GJ_0t}{c^2\rho^3} +\tan^{-1} [\cos i\tan (\Omega t+\beta )]+\varphi_0,
\end{equation}
where, $\varphi_0$ is an integration constant and $i$ is the inclination angle.

As a result we can state that the above considered spherical orbits can be characterized as circular orbits in the post-Newtonian gravitational field of $\Lambda$-gravity, which undergo LT precession due to the presence of a constant angular momentum of the source. 
Consequently, we can find the modified Keplerian frequency
\begin{equation}
\omega_{\Lambda} = \frac{GM}{r^3} - \frac{\Lambda c^2}{3}
\end{equation}
and the frequency $\omega$ as follows
\begin{equation}
\Omega = \omega - \frac{3GJ_0}{c^2\rho^3}C,
\end{equation}
the form for $\omega$ is obtained from the relation E.(\ref{Om2})
\begin{equation*}
\omega = \frac{GM}{\rho^3}\left(1-\frac{3GM}{c^2\rho} - \frac{\Lambda\rho^2}{3}\right) - \frac{\Lambda c^2}{3}\left(1 + \frac{GM}{c^2\rho} - \frac{\Lambda\rho^2}{3}\right).
\end{equation*}
Thus, in comparison with the results of \cite{M1}, both definitions of constants $\Omega$ and $\omega$ are changed accordingly. Namely, for both of them we get relativistic and non-relativistic corrections due to the presence of $\Lambda$ term. However, since these corrections enter into the equations of motion as a combination of constants, we can conclude that apart from the numerical corrections, the nature of the analysis regarding the spherical orbits does not change.

\section{Perturbed orbits}

Turning to the perturbation analysis, we can state that in case of 
$$J = J_0 + J_1t$$ 
the perturbation $(f(t), g(t),h(t))$ which is written as
\begin{equation}\label{pert}
r=\rho (1+f(t)),\quad \theta =\vartheta+g(t),\quad \phi=\varphi +h(t),
\end{equation}
will lead to the instability if the spherical orbits, causing inward (when $J_1 \cos i<0$) and outward (when $J_1 \cos i>0$) spiral orbits. 

Considering the physics of orbital mechanics and the possible modifications of pure Keplerian motion, it is essential to mention the relativistic shift of the precession, too. Namely, it is a deviation from the precession predicted by Newtonian gravity which occurs due to GR effects and is equal to

\begin{equation}\label{Shift}
\Delta \phi _{GR} = 6\pi \frac{GM}{c^2 a (1-e^2)},
\end{equation}
where $a$ and $e$ are the semi-major axis and the eccentricity of the orbit, respectively. Accordingly, if we consider the $\Lambda$-term in the equations, besides the $\Delta \phi_{GR}$ we get an additional shift
\begin{equation}\label{ShiftL}
\Delta \phi _{\Lambda} = \frac{\pi c^2 \Lambda a^3}{GM} \sqrt{1-e^2}.
\end{equation}
It can be shown that, this additional shift is of pure relativistic origin and cannot be reduced to any non-relativistic effect \cite{Br,Mash}. Moreover, in the absence of central object of mass $M$, the $\Lambda$-term in Eqs.(\ref{mod}),(\ref{GEMmetric}) will have no effect on the orbital precession, and, of course, for real astrophysical systems the $\Delta \phi _{\Lambda}$ numerically is essentially smaller than $\Delta \phi_{GR}$.

Finally, it is worth mentioning that besides the above derivation, we can have another relation where right-hand side of Eq.(\ref{int}) vanishes  generally,
\begin{equation}\label{crit}
r^3_{crit} = \frac{3 GM}{c^2 \Lambda},
\end{equation}
which is the distance where the gravitational repulsion of $\Lambda$-term in Eq.(\ref{mod}) completely balances the gravitational attraction of Newtonian term. It should be stressed that although in this case, the test particle located at $r_{crit}$ does not feel any force from the central object it cannot be considered as a free particle. Indeed, this is a unique feature which is taken place in the context of $\Lambda$-gravity. The reason lies on the fact that while the net force over the test particle with mass $m$ vanishes, the gravitational potential is non-zero i.e.
\begin{equation}
\Phi = -\frac{GM}{r} - \frac{\Lambda c^2 r^2}{6}|_{r_{crit}}=- \frac{\Lambda c^2 r_{crit}^2}{2}, \quad   F =-m\nabla \Phi = -\frac{GMm}{r^2} + \frac{\Lambda c^2 rm}{3}|_{r_{crit}}=0.
\end{equation}

\section{GEM interpretation}

In this section we intend to generalize the standard ``GEM" interpretation in such way that it incorporates the $\Lambda$ term. Namely, the GEM is an approaches which tries to find an analogy between the Maxwell field equations and those of GR written at relevant approximation \cite{GEM}. Thus, in the context of $\Lambda$-gravity the GEM equations will be written as
\begin{equation}\label{EL}
\nabla \cdot \mathbf{E} =  4 \pi G \rho - \Lambda c^2,
\end{equation}
\begin{equation}
\nabla \times \mathbf{E} = - \frac{\partial \mathbf{B}}{\partial t},
\end{equation}
\begin{equation}
\nabla \cdot \mathbf{B} = 0,
\end{equation}
\begin{equation}
\nabla \times \mathbf{B} = \mu \mathbf{J} + \frac{\partial \mathbf{E}}{\partial t},
\end{equation}
where $\Lambda$-term in the first equation can be regarded as the additional ``vacuum density" which is equal to $\frac{-\Lambda c^2}{4 \pi G}$. Considering the fact that the vacuum density is just a combination of constants, we can state that it won't change anything in the continuity equation
\begin{equation}
\frac{d \rho}{d t} + \nabla \cdot \mathbf{J} = 0.
\end{equation}
Meantime, in the context of GEM the LT precession is interpreted as the Larmor precession which is written as
\begin{equation}
\Omega = \gamma \mathbf{B},
\end{equation}  
where $\gamma$  is the gyromagnetic ratio and $\mathbf{B}$ is the magnetic field. Accordingly, for LT precession, the gravitomagnetic field will be written as
\begin{equation}
\quad \mathbf{B} = \frac{G \mathbf{J}}{c^2 r^3},
\end{equation}
By turning to $\Lambda$-gravity it can be shown that the LT precession is written as 
\begin{equation}\label{LTL}
\frac{2 G J}{c^2 r^3} + \frac{\Lambda J}{3 M}.
\end{equation}
Considering the fact that, the gravitomagnetic field is produced by gravito-current i.e. the rotation, it is not correct to consider the $\frac{\Lambda J}{3M}$ as a correction to the magnetic field. Here the key point is that the second term in Eq.(\ref{LTL}) can be interpreted as the correction over the gravito-gyromagnetic ratio
\begin{equation}\label{GyrL}
\gamma = 2 \left(1 + \frac{M_{\Lambda}}{M}\right),
\end{equation}
where $M_{\Lambda}$ is the ``effective mass" of the vacuum with a density $\frac{\Lambda c^2}{8 \pi G}$. It can be checked that this correction is purely relativistic in its nature. Namely, comparing both Eq.(\ref{mod}) and Eq.(\ref{EL}), it becomes clear that while the presence of $\Lambda$ in the relativistic equations is regarded as the contribution of vacuum with density equal to $\frac{\Lambda c^2}{8 \pi G}$, in the non-relativistic limit, where we are dealing with gravitational force, this correction changes to the notion of vacuum with density equal to $\frac{- \Lambda c^2}{4 \pi G}$. Speaking in other words, one can conclude that while the $\Lambda$ term can contribute to the gravitational potential as an additional term, for those equations where instead of potential we are dealing with the notion of ``gravitational fields" or ``gravitational force" the contribution of $\Lambda$ becomes subtractive. Thus, by checking the presence of $\Lambda$ term in Eq.(\ref{LTL}) we can state that this correction should be a relativistic correction with no classical analogue. Considering the Eq.(\ref{GyrL}) and continuing the analogue in electromagnetism in the GEM spirit, we can state that such correction is similar to the electron's g-factor $g_e$ in the context of relativistic quantum mechanics
\begin{equation}\label{ge}
g_e = 2(1+\frac{\alpha}{2 \pi}+...),\quad \alpha = \frac{1}{4 \pi \epsilon}\frac{e^2}{\hbar c} \approx \frac{1}{137}.
\end{equation}
However, the main difference between Eq(\ref{GyrL}) and above relation is that while in Eq.(\ref{ge}) the second term is constant, the $\frac{M_{\Lambda}}{M}$ is variable. Namely, its value becomes larger as the radius increases. Moreover, it can be shown at some distance its contribution will be equal to the first term i.e.
\begin{equation}\label{LTcrit}
r^3 = \frac{6 GM}{c^2 \Lambda},
\end{equation} 
which is twice the $r_{crit}$ in Eq.(\ref{crit}). Consequently, we can state that at radii larger than $r$ in Eq.(\ref{LTcrit}) the dominant term such causes the LT precession is the $\Lambda$ term. However, it should be recalled that even in such cases, the presence of a central rotating body with mass $M$ and angular momentum $J$ is the necessary condition. Indeed, this statement once again shows that the nature of $\Lambda$-term is purely relativistic and confirms the fact the in contrast to other similar precessions e.g. the geodetic precession, for LT precession the rotation of the central object is essential.

\section{Conclusions}

In this paper we studied the geodesics of spherical bound orbits, i.e. the circular orbits with frame dragging, in the context of $\Lambda$-gravity within the gravito-electromagnetic formalism. We have derived the equations in the same analogy with the equations of motion for a charged particle experiencing the Lorenz force. Consequently, we have shown that the $\Lambda$-term enters both the relativistic as well as non-relativistic corrections to the original equations. Furthermore, by considering the LT precession, we have given a new interpretation of GEM within which the $\Lambda$ is included. Namely, we have shown that, from the GEM point of view, the additional $\Lambda$ term in LT can be regarded as a relativistic correction to the gravito-gyromagnetic ratio.

\end{document}